\documentclass[preprint,showpacs,preprintnumbers,amsmath,amssymb]{revtex4}
\usepackage{graphicx}
\usepackage{dcolumn}
\usepackage{bm}
\begin{document}
\preprint{preprint - vortex group, tifr}
\title{Manifestation of modulation in spatial order of flux line lattice in weakly pinned superconductors : a vibrating sample magnetometer study}
\author{A. D. Thakur$^1$ \footnote{Electronic address: ajay@tifr.res.in}, D. Pal$^2$, M. J. Higgins$^3$, S. Ramakrishnan$^1$, A. K. Grover$^1$ \footnote{Electronic address: grover@tifr.res.in} }
\affiliation{$^1$ Department of Condensed Matter Physics and Materials Science, Tata Institute of Fundamental Research, Mumbai 400005, India \\
$^2$ Department of Physics, University of Notre Dame, Notre Dame, IL 46556, USA \\
$^3$ Department of Physics and Astronomy, Rutgers University, Picastaway, New Jersey 08540, USA}

\date{\today}

\begin{abstract}
We have investigated the healing of transient disordered vortex states injected into a superconducting sample during the field ramping process in isothermal M-H measurements as a prescription to comprehend the modulation in the state of order of the underlying equilibrium vortex state in weakly pinned single crystals of $2H$-NbSe$_2$ (zero field transition temperature, $T_c(0)~\sim~6~K$) and Ca$_3$Rh$_4$Sn$_{13}$ ($T_c(0)~\sim~8.2~K$), which display order-disorder transition(s). Our results reaffirm the notion that spatial order in the flux line lattice proceeds to the amorphous state in two distinct steps as evidenced by the presence of second magnetization peak (SMP) anomaly and the peak effect (PE) phenomenon. Disordering commencing at the onset field of SMP has a window to heal between its peak field ($H_{smp}^p$) and the onset field ($H_p^{on}$) of the PE. A comparison of the scan rate dependence of magnetization hysteresis widths in two different crystals of $2H$-NbSe$_2$, one of which displays only the PE phenomenon and the other one that displays both SMP and PE anomalies, show that while enhancment in quenched random pinning invokes the occurence of SMP anomaly, it slows down the temporal decay of currents across the PE region.
\end{abstract}
\maketitle

\section{Introduction}
Evidences have accumulated in recent years that the peak effect (PE) phenomenon in critical current density ($J_c$) located at the edge of the depinning process is distinct from the second magnetization peak (SMP) anomaly commencing at field values deeper in the mixed state of the weakly pinned superconducting samples of several low $T_c$ and high $T_c$ compounds, e.g., Ca$_3$Rh$_4$Sn$_{13}$ \cite{sarkarpal}, RNi$_2$B$_2$C ($R=Y,~Lu$) \cite{djn}, V$_3$Si \cite{kupfer}, $2H$-NbSe$_2$ \cite{adtprb, adtpramana}, PrOs$_4$Sb$_{12}$ \cite{kobayashi}, YBa$_2$Cu$_3$O$_7$ \cite{sarkarpal, degroot, dpal, nishizaki}, etc. In collective pinning \cite{lo74} scenarios, the critical current density can be related to the length scales over which the displacements of the flux lines from their equilibrium positions remain correlated. In such frameworks, an anomalous increase in field/temperature (H/T) variation of $J_c$ can be construed to imply a decrease in the correlation volume over which flux lines remain elastically deformed and collectively pinned, thereby, leading to the notion of order-disorder transformation in the underlying flux line lattice (FLL) \cite{lo79, koopman}. The occurrence of two anomalous variations in $J_c(H)$ (e.g., the SMP and the PE) in a given isothermal scan may, therefore, imply that the disordering of the ordered lattice can proceed in two distinct steps, before reaching the depinning limit in the bulk of the sample \cite{adtprb, adtjpsj}. A notion of weak pinning by surface inhomogeneities may survive, after crossing the bulk depinning limit, upto the upper critical field ($H_{c2}$) \cite{pautrat}.

Recent studies of the metastability effects \cite{adtprb} in single crystals of $2H$-NbSe$_2$, which display a SMP like anomaly prior to PE, have revealed that ordered and disordered regions co-exist from above the onset position of SMP anomaly upto a limiting temperature, designated as the spinodal temperature ($T^{\star}$), which lies above the peak position of PE \cite{zlxiao}. A spinodal line is the locus of $T^{\star}(H)$ above which the path dependence in $J_c(H, T)$ is absent \cite{adtprb}. The notion of spinodal line is rationalized by terming the order-disorder transition in FLL of $2H$-NbSe$_2$ system to have a first order character, in analogy to the FLL melting line in the weakly pinned crystals of high $T_c$ bismuth cuprates \cite{khaykovich}. Li and Rosenstein \cite{lirosenstein1, lirosenstein2} have theroetically argued that metastability associated with the order-disorder first order transition line could span a large part of the ($H,~T$) phase space, starting from the spinodal line and extending sufficiently below the boundary marking the onset position of the order-disorder transition. Using fast current ramp procedures in transport measurements, Xiao {\it et al} \cite{zlxiao} had exemplified the notion of a generic spinodal line via a normalized field-temperature ($t^{\star}=T^{\star}/T_c(0), h^{\star}=H^{\star}/H_{c2}(0)$) plots in a number of crystals of $2H$-NbSe$_2$, which are in agreement with the theoretical results of Li-Rosenstein \cite{lirosenstein1, lirosenstein2}. The single crystal samples studied by Xiao {\it et al} \cite{zlxiao} displayed only the PE phenomenon. Extending these studies, Li {\it et al} \cite{gli} have recently exploited the time resolved transport measurements in a high purity single crystal of $2H$-NbSe$_2$ ($T_c(0) \sim 7.2~K$) to elucidate the marking of a limiting temperature $T_L(H)$ far below the onset temperature of PE ($T_p^{on}(H)$), below which (i.e., $T<T_L(H)$), FLL is essentially dislocation free. Above $T_L(H)$, a sharp change in the reordering timescales, observation of strong hysteresis, and the presence of a plastic response, reflects the proliferation of dislocations, which they surmise as a precursor of the PE transition to the disordered state. These results in turn imply that in the (H, T) phase-space between the onset of precursor effects and upto the spinodal line (i.e., between $T_L(H)$ and $T^{\star}$ lines), the manifestation of tendency to polycrystallinity in the vortex matter can be explored in a variety of vortex dynamics measurements.

Explorations by Banerjee {\it et al} \cite{ssbphysica} and Thakur {\it et al} \cite{adtprb, adtpramana} in a variety of crystals of $2H$-NbSe$_2$ reveal an intimate connection between hysteretic effects (i.e., metastability/thermomagnetic response) and effective pinning, such that an increase in quenched random disorder leads to surfacing of a SMP anomaly prior to the PE phenomenon. The results of Li {\it et al} \cite{gli} have motivated us to explore a manifestation of the evolution in the reordering time scales when the notion of phase co-existence \cite{marchevsky} spans across the regions of SMP and PE anomalies \cite{adtprb, adtpramana}, and it can be bifurcated into two sub-parts, where ordered and disordered pockets dominate, respectively. In particular, in the sub-part between the onset positions of SMP and PE, where the ordered pockets dominate, there is a possibility of further sub-division. While the increase in $J_c$ from the onset position of SMP ($H_{smp}^{on}$) anomaly to its peak position ($H_{smp}^p$) can be termed as anomalous, the decrease in $J_c$ from $H_{smp}^p$ to the onset position of PE ($H_{p}^{on}$) is indeed non-anomalous \cite{adtprb, adtpramana}. This permits a non-monotonic variation in the relative weights of ordered and disordered pockets in the phase co-existence regime between $H_{smp}^{on}$ and $H_p^{on}$. Such a behavior could get reflected in the reordering time scales experienced during isothermal magnetization hysteresis experiments, where the magnetic field is steadily ramped as in a Vibrating Sample Magnetometer (VSM).

In field ramping experiments, disordered vortices enter a given sample through its  edges/surface inhomogeneities \cite{paltedgeeffect} and they attempt to reach towards an underlying equilibrium state in specific experimental conditions \cite{giller, beek, kalisky1, kalisky2, kalisky3}. We present here an analysis of the results of one such study performed using a VSM. We believe that new data are supportive of a claim \cite{adtprb} that two distinct features (viz., the SMP and PE anomalies) evident in the disordering of the weakly pinned FLL represent two distinct phase transitions. 

\section{Experimental}

The single crystals of $2H$-NbSe$_2$ used in this study are, (i) a nascent pinned sample $X$ ($4\times1.74\times0.18~mm^{3}$) with $T_c(0) \sim 7.22~K$ and  (ii) few somewhat more strongly pinned samples, viz., $Z$ ($3.25\times2\times0.14~mm^{3}$) and $Z^{\prime}$ ($2.5\times1.5\times0.14~mm^{3}$), having 200~ppm of Fe as impurity and with $T_c(0) \sim 6~K$. The residual resistivity ratio ($\frac{R_{300~K}}{R_{8~K}}$) for the samples $X$ and $Z^{\prime}$ were measured to be 20 and 9.8, respectively. The crystal of Ca$_3$Rh$_4$Sn$_{13}$ ($T_c(0)~\sim~8.2~K$) is the same \cite{tomy} as utilized by Sarkar {\it et al} \cite{sarkarpal}. Isothermal M-H loops were recorded on a 12~Tesla Vibrating Sample Magnetometer (VSM) (Oxford Instruments, UK), with field scan rates ($\dot{H} = \frac{dH}{dt}$) in the range of 100~Oe/min to 8~kOe/min. The ac susceptibility measurements were done utilizing an ac option facility on a commercial SQUID magnetometer (Quantum Design Inc., U.S.A., Model MPMS7).

\section{Results and Discussion}

\subsection{Isothermal M-H measurements and scan rate dependence of hysteresis width in crystal $Z^{\prime}$ of $2H$-NbSe$_2$}

Fig. 1 (a) shows a portion of an isothermal M-H loop recorded at 3.86~K in crystal $Z^{\prime}$ of $2H$-NbSe$_2$, with the magnetic field having being ramped with $\dot{H}$ of 3~kOe/min. The PE and SMP like anomaly have been identified, with the boxed region showing the latter, which sits deeper in the mixed state. The same figure also depicts magnetization values (open triangles) at different fields obtained in field cooled (FC) manner ($M_{FC}$). In view of the tendancy that notion of pinning resists the expulsion of flux from the interior of the sample during field cooling process, the small angle neutron scattering experiments to probe the structure of the vortex matter at a given field are performed in the field cooled conditions \cite{pautrat}. In a FC state, the magnetization value can, therefore, be taken to represent a notional equilibrium magnetization value ($M_{eq}$), with no gradient in macroscopic field in the bulk of the sample. The width of the magnetization hysteresis loop ($\Delta M(H)$) at a given field is usually considered to measure the critical current density $J_c(H)$ \cite{fietzandwebb}. The hysteresis width is observed to depend on the scan rate $\dot{H}$ of the field. The linear relationship between $\Delta M(H)$ and $J_c(H)$ is satisfactory when field fully penetrates the sample and its variation over the sample is not large. In the given crystal $Z^{\prime}$, the full penetration field at 3.86~K in the field range of interest is less than 100~Oe \cite{adtthesis}. We may, therefore, assume that in the field interval of interest (1~kOe to 15~kOe) in the present experiment in sample $Z^{\prime}$ at 3.86~K, the variation in the local macroscopic field across the sample is a very small fraction of the externally applied field. If we, therefore, assume that the applied field value (notionally) prevails across the sample, it will amount to a little quantitative inaccuracy.

In Fig. 1 (b), we have shown the effect of $\dot{H}$ on one-half of the hysteresis width, $\Delta M^{for}(H)$ ($= -(M^{for}(H) - M_{FC}(H))$ and proportional to the critical current density, $J_c^{for}(H)$, during the forward leg of the envelope curve from 1~kOe to 15~kOe. These data suggest a substantially larger scan rate dependence of the $J_c^{for}(H)$ values prior to the PE region. It is apparent that $J_c(H)$ at any H decreases as ramp rate decreases. An instantaneous value of $J_c(H)$ would correspond to the limit of an infinite scan rate, i.e., $(\dot{H})^{-1}~\rightarrow~0$. Slower ramping of field permits larger settling time for injected disordered bundles of vortices to approach towards an underlying equilibrium state at a given $H$. In order to elucidate the modulation in the field dependence of $J_c(H)$ vs $\dot{H}$ before and after the order-disorder transition(s) as evidenced by SMP and PE, we plotted $\Delta M^{for}$ vs $\dot{H}^{-1}$ at different values of $H$. From every such plot, we extracted by extrapolation a limiting magnetization value ($\Delta M^{for, \infty}$) for an infinite scan rate ($\dot{H}^{-1} \rightarrow 0$). Using these values, we show in the panels (a) and (b) of Fig. 2 the log-log plots of $\Delta M^{for}/ \Delta M^{for,~\infty}$ (proportional to the normalized critical current density, $J_c^{for, norm}$) versus $\dot{H}^{-1}$ at different field values, starting from deep within the mixed state, right upto the peak position of the PE at 3.86~K. From these data, it is evident (see Fig. 2 (a)) that the temporal decay of the forward critical current density progressively increase as one proceeds from the low field end (e.g., 1.6~kOe), till one approaches the onset field of the SMP, $H_{smp}^{on}$ ($\approx 6.9~kOe$). Above that field, there is a relative decrease in the temporal decay, until the peak position of SMP, $H_{smp}^p$ ($\approx 7.9~kOe$) is reached. Fig. 2 (b) shows that from $H_{smp}^p$ to the onset field for the PE, $H_p^{on}$ ($\approx 9.0~kOe$), there is again an increase in the temporal decay. Beyond $H_p^{on}$, the temporal decay reduces monotonically upto the peak position of the PE, $H_p$ ($\approx 12.5~kOe$). These observations, we believe, attest to the presence of multiple undulations in the reordering time scales of the injected disordered vortices. 

\subsection{Modulation in the normalized decay rate across the phase co-existence regime}

In order to further comprehend the modulation in the temporal decays in $J_c^{for, norm}$, we have used a parameter $S$ defined as $S=\mid\frac{d~log(J_c^{for})}{d~log(\dot{H}^{-1})}\mid _{\dot{H}^{-1}~\rightarrow~\infty}$, which we believe, represents the normalized decay rate of $J_c$ at larger times (i.e., very slow scan rate). The inset and main panels of Fig. 3 show behavior of $S~vs~H$ at 3.86~K in the sample $Z^{\prime}$ of $2H$-NbSe$_2$. We indeed observe a non-monotonic variation in $S(H)$, until it reaches a limit near $H=H_p$, which we reckon correspond to a value that such a parameter would assume for the usual thermal creep behavior \cite{kimanderson, yeshurunrmp} in the pinned state.

As stated earlier, the observed behavior can be rationalized on the premise that if the underlying equilibrium state at a given H is completely disordered, the injected disordered bundles of vortices would assume such a state instantaneously and , therefore, the time decay (i.e., scan rate dependence) in hysteresis width ($\propto~J_c(H)$) would be governed by the process of thermal creep alone. However, if the underlying equilibrium state is partially ordered, the injected vortices would require additional time lapse to conform to that state before the process of thermal creep starts governing the observed time decay. In an isothermal ramp, at low fields (where FLL lattice constant $a_0 > \lambda$, the penetration depth), the injected diordered vortices could be expected to first conform to the reentrant disordered state \cite{ssbppf}, followed by (gradual) crossover to the progressively better ordered elastic glass phase \cite{gialed} upto the onset of the SMP anomaly. Thereafter, as the state of co-existence of weaker and stronger pinned phases commences \cite{marchevsky, adtprb}, the injected disorder would attempt to heal towards the partially ordered equilibrium state, which can be attained in somewhat shorter time scales than that needed for $H \leq H_{smp}^{on}$. We believe that the modulation in $S(H)$ in Fig. 3 essentially reflect the changes in the underlying equilibrium state from reentrant disordered phase to the peak position of PE. We are tempted to attribute the turnaround in $S(H)$ response between $H_{smp}^p$ and $H_p^{on}$ to the possibility of improvement in the state of partial order with increase in field in this interval \cite{dpalsst}. Considering that the $S(H)$ value at $H_p^{on}$ is about the same as that at $H_{smp}^{on}$, we may naively state that the partially disordered phase nearly heals uptill the arrival of the PE regime.

We show in Fig. 4 the scan rate dependence of the ratios of the critical current density at the peak position to that at the onset position for the SMP and PE anomalies, respectively obtained from the isothermal M-H data at 3.86~K. The increase in the values of the ratio with decrease in scan rate conforms to the notion that slower ramp rate permits attainment of superior spatial  order prior to entry into the anomalous regimes. For the SMP anomaly, the ratio changes from about 1.3 to about 2.5, whereas for the PE, it changes from about 20 to about 90. Let us recall that in a collective pinning scenario due to Larkin-Ovchinnikov \cite{lo74}, $J_c$ relates inversely to the correlation volume, $V_c$ ($J_c \propto 1/ \sqrt{V_c} \propto 1/ \sqrt{R_c^{2}L_c}$, where $R_c$ and $L_c$ are radial and longitudinal correlation lengths; for simplicity we may take $L_c \propto R_c$). The above values then imply a reduction in radial correlation length ($R_c$) by a factor of about 20 across the PE. Banerjee {\it et al} \cite{ssbphysica} had estimated $R_c / a_0$ (where, $a_0$ is FLL constant) at $H=10~kOe(\|c)$ to be $\sim$20 at 4.2~K in a crystal (sample $Z$) of $2H$-NbSe$_2$ ($T_c(0) \sim 6~K$). A change in $R_c$ value by a factor of about 20 during complete amorphization of FLL between the onset and peak positions of PE therefore appears satisfactory. When both SMP and PE are present in a given isothermal scan, the collapse in correlation volume of vortex lattice could occur in two stages - a very large domain first sub-divides into multiple parts, followed by partial healing and (eventual) complete amorphization into micro-domains.

To establish the efficacy of prescription of analysis of scan rate dependence of hysteresis width as illustrated in Fig. 2 and Fig. 3., we show for comparison in Fig. 5 the $S(H)$ data at two temperatures in a weakly pinned crystal of another low $T_c$ superconductor, viz., Ca$_3$Rh$_4$Sn$_{13}$, in which SMP and PE had been reported earlier \cite{sarkarpal}. The SMP anomaly in this compound is as prominent as the PE, the field region of SMP, however, is insensitive to the temperature variation, whereas the PE closely follows the $H_{c2}$ line. The separation between the SMP anomaly and the PE, therefore, decreases as $T$ increases. It had also been suggested that in Ca$_3$Rh$_4$Sn$_{13}$, the dislocations injected in between the onset and peak positions of SMP could heal upto the onset position of the PE \cite{dpalsst}. It is useful to examine the $S(H)$ behavior in Ca$_3$Rh$_4$Sn$_{13}$ at 1.9~K and 3.6~K (cf. Fig. 5 (a) and 5 (b)). While the non-monotonic behavior in $S(H)$ is evident at both temperatures, the relative enhancement in $S(H)$ values between $H_{smp}^{on}$ and $H_{smp}^p$ is more at 1.9~K as compared to that at 3.6~K. Considering the larger field interval between $H_{smp}^{on}$ and $H_{smp}^p$ in Ca$_3$Rh$_4$Sn$_{13}$ as compared to the corresponding interval in $2H$-NbSe$_2$, the $S(H)$ values decrease to a much deeper level in the former sample, and from which the turnaround also happens upto a lower limit, i.e., $S(H_p^{on})$ value is smaller than $S(H_{smp}^{on})$ value in Ca$_3$Rh$_4$Sn$_{13}$.

\subsection{Effect of variation in pinning on the scan rate dependence of hysteresis width across SMP and PE anomalies in crystals of $2H$-NbSe$_2$}

A comparative look at $S(H)$ data in crystal $Z^{\prime}$ of $2H$-NbSe$_2$ and that of Ca$_3$Rh$_4$Sn$_{13}$ conveys similarity in behavior across SMP and PE anomalies (cf. Fig. 3 and Fig. 5), however, the values of the $S$ parameter in the former case are much larger. The observation that SMP anomaly in Ca$_3$Rh$_4$Sn$_{13}$ does not display temperature variation is supportive of the viewpoint that this anomaly is entirely pinning induced. A study of the variation of quenched random pinning on the PE phenomenon in a number of crystals of $2H$-NbSe$_2$ had independently revealed that SMP anomaly in this system surfaces up with the enhancement of pinning. It is, therefore, instructive to compare the scan rate dependences of hysteresis width in crystals of $2H$-NbSe$_2$, with varying pinning strengths. In a given crystal of $2H$-NbSe$_2$, effective pinning can also be seen to enhance with the increase in field \cite{worthington, menondasgupta, vinokur}.

We present in Fig. 6 glimpses into few above stated results obtained in crystals $Z^{\prime}$ and $Z$ (both having $T_c(0) \sim 6~K$) and a nascent pinned crystal $X$ of $2H$-NbSe$_2$ ($T_c(0) \sim 7.2~K$). Fig. 6 (a) shows plots of critical current density on the forward leg normalized to its value at $H \sim 0~Oe$, $J_c^{for}$ ($\propto \Delta M^{for} / \Delta M^{for} (H \sim 0~kOe)$), at scan rates of 3~kOe/min and 0.3~kOe/min, respectively, at a temperature of 3.07~K in the crystal $Z^{\prime}$ of $2H$-NbSe$_2$. These data correspond to a situation, where the SMP anomaly partially overlaps with the PE anomaly (broadening due to enhancement in effective disorder) and one can barely mark the onset position $H_p^{on}$ of the PE. Even in such a circumstance, one can see a marked difference in the dynamical response above the onset field of the PE anomaly and at field values below it. One may be tempted to ignore the distinctiveness of the region of SMP anomaly on the basis of data in Fig. 6 (a). An inset panel in Fig. 6(a) shows the ramp rate dependence of normalized $\Delta M(H)$ at 2~K in the crystal $Z$ of $2H$-NbSe$_2$. Here, one can notice that at a slower scan rate of 0.25~kOe/min., the signature corresponding to the SMP like anomaly becomes indiscernible, while it is distinctly observable at a larger scan rate of 8~kOe/min. Such an observation could lead to an apprehension whether any change occurs in the underlying vortex state across the field interval of $H_{smp}^{on}$ and $H_{smp}^p$. In this context, we now draw attention to the scan rate dependence of the hysteresis width across PE in crystal $X$ (see Fig. 6 (b)). An inset panel in Fig.~6(b) shows a comparison of the forward legs of two M-H loops obtained with $\dot{H} = 5~kOe/min$ and $0.5~kOe/min$, respectively in crystal $X$ at 2~K. It is apparent that the hysteresis width across PE region shrinks rapidly with decrease in $\dot{H}$. Attempts to obtain a hysteresis bubble in the M-H data across PE recorded using a SQUID magnetometer (Quantum Design Inc. U.S.A.), where data are recorded in a quasi-static manner (after a time lapse of about a minute from the instance the magnet current is set in persistent mode) did not yield affirmative output, as the hysteresis bubble had collapsed to the background limit in the elapsed time \cite{troyanovsky, ssbthesis}. Non-observation of PE bubble in M-H data in very clean crystals of $2H$-NbSe$_2$ has been eluded to earlier \cite{troyanovsky}, however, the changes (amounting to a phase transition) happening in the vortex state across the PE have been elucidated via STM imaging in the same crystal \cite{troyanovsky}. We are, therefore, inclined to persue probing the field interval of SMP region in the crystal $Z^{\prime}$ of $2H$-NbSe$_2$.

In Fig.~7, we show plots of the widths of hysteresis bubble at $H=H_{smp}^p$ and $H=H_p$ normalized to its extrapolated value at  $(\dot{H})^{-1} \rightarrow 0$, as a function of $(\dot{H})^{-1}$ at 3.86~K in crystal $Z^{\prime}$. We have also included in this figure the plot of the hysteresis width at $H=H_p$ in crystal $X$ at 2~K. Note the similarity in the behavior of plot pertaining to PE in crystal $X$ with that pertaining to SMP in crystal $Z^{\prime}$. Both the hysteresis widths rapidly decrease with increase in $(\dot{H})^{-1}$, thereby, implying the difficulty in their observation in the static limit while recording data with a SQUID magnetometer \cite{ssbthesis}. The width of the PE bubble in crystal $Z^{\prime}$, however, decreases slowly with increase in $(\dot{H})^{-1}$ and this anomaly, therefore, registers its presence in the data recorded with the SQUID magnetometer. The observations in Fig.~6 and Fig.~7 prompt us to state that enhancement in pining in crystals of $2H$-NbSe$_2$ not only invokes the surfacing of SMP, but, also slows down the time decay of macroscopic currents that get set up in response to a driving force in the PE region.

\subsection{Effect of frequency and amplitude of ac field on SMP and PE in crystal $Z^{\prime}$ of $2H$-NbSe$_2$ and construction of its vortex phase diagram}

One difference between SMP and PE anomalies in Ca$_3$Rh$_4$Sn$_{13}$ and the sample $Z^{\prime}$ of $2H$-NbSe$_2$ is that in the former both the features are well separated at lower temperatures (higher fields), but in the latter sample, SMP appears  as a precursor part of PE, with the possibility of large overlap at lower temperatures. When SMP and PE lie in closer proximity, the two together present themselves as a candidate \cite{xiaoedgeeffect} of edge effects complicating the PE pertaining to order-disorder transition in the bulk of the sample in ac susceptibility measurements. In order to further explore the case of SMP as a bulk feature and/or edge effects, we measured ac susceptibility in sample $Z^{\prime}$ of $2H$-NbSe$_2$ at different frequencies and amplitudes of ac field. 

Fig.~8 (a) shows isofield ($H=10~kOe,~\|c$) ac susceptibility ($\chi^{\prime}_{ac}(T)$) data measured with an $h_{ac}$ of 2.5~Oe (r.m.s.) in the frequency range 1~Hz to 1111~Hz. It can be noted that the signatures corresponding to SMP and PE anomalies remain distinctly observable over the entire range of frequencies, thereby discounting the possibility of SMP anomaly as being an edge effect. Fig.~8 (b) shows the $\chi^{\prime}_{ac}(T)$ data measured with different amplitudes of the ac drive ($h_{ac}$) at a frequency of 111~Hz. The omnipresence of the two anomalous variations is self evident in Fig.~8. Change of frequency by three orders ($\sim 1-10^3~Hz$) is expected to compromise the edge effects in a preferential manner \cite{paltedgeeffect}, however, this does not seem to be the case in the present data. We surmise that SMP and PE represent occurrence of two distinct order-disorder transitions in (H, T) phase space, whose locations can be marked from appropriate sets of data. Fig. 9 shows the vortex phase diagram for H$\|$c in crystal $Z^{\prime}$ of $2H$-NbSe$_2$, sketched from M-H data recorded at a high scan rate (3~kOe/min). The reentrant disordered phase, the Bragg glass (BG) phase, the phase co-existence region (Vortex glass (VG) phase) and the surface pinning region have been identified. The bifurcation of the phase co-existence region into regions I and II, where ordered and disordered pockets of vortices dominate, respectively, is also shown. This diagram covers a larger portion of (H, T) space as compared to the one presented earlier \cite{adtprb, adtpramana} in this sample. It is appropriate to add here a caveat that phase boundaries in Fig. 9 have been drawn from bulk magnetization studies, these would get somewhat modified, when drawn using local magnetization measurements.

The construction of the vortex phase diagram in Fig. 9 presumes the applicability of notions of Bragg glass to vortex glass transition and an inverse relationship between the critical current density and the spatial extent over which the vortices remain correlated in the spirit of Larkin-Ovchinnikov collective pinning theory \cite{lo74, lo79}. By combining electrical transport studies in strip and Corbino geometry with Bitter decoration patterns at low fields (viz., 36~Oe, where intervortex spacing $a_0 >$~penetration depth $\lambda$) in $2H$-NbSe$_2$ for $H~\|~c$, it has been surmised \cite{fasano, menghini} that no obvious correlation exists between the topology of the vortex structure and enhancement in critical current in the (partially) ordered (i.e., polycrystalline) vortex matter. An extension of the notion of absence of any correspondence between the critical current density and the spatial order in the vortex state in the field region where interaction effects between vortices are appreciable ($a_0 \ll \lambda$) could, in principle, undermine the viewpoint pursued above.

\section{Summary and Conclusion}

We have presented experimental data from dc magnetization measurements to elucidate the notion of a modulation in the amount of the ordered (disordered) fraction in the underlying equilibrium/steady state in a field ramp up cycle in weakly pinned samples showing second magnetization peak anomaly and peak effect phenomenon. The annealing to an underlying equilibrium/steady state of the transient disordered vortex bundles injected into the sample during the field ramp in a VSM measurement has been used as a prescription to gain information on the extent of order/disorder prevailing in the vortex state at a given (H, T). From an analysis of the ramp rate dependence of magnetization hysteresis width, a parameter $S$ ($=\mid\frac{d~log(J_c^{for})}{d~log(\dot{H}^{-1})}\mid _{\dot{H}^{-1}~\rightarrow~\infty}$) has been projected, which purports to correlate to the time required to assume an underlying equilibrium state. Observation of non-monotonic variation in $S(H)$ can be understood within a premise that the extent of spatial order in the equilibrium/steady state of the vortex matter at a given $H$ determines the (initial) scan rate dependence of the hysteretic magnetization response. The magnetization value at a given instance at a particular field during a field ramp cycle carries information on the incomplete annealing of the stronger pinned disordered pockets towards the underlying equilibrium/steady state \cite{giller, beek, kalisky1, kalisky2, kalisky3}. When the underlying equilibrium/steady state is anticipated to have a high degree of order, the injected disordered bundles require longer time lapse and resultant $S(H)$ value is large. On the other extreme, when the destined equilibrium/steady state is disordered in nature, the injected disordered bundles assume that state instantly and S parameter is small (its value corresponding to flux creep alone). At the low field end, where vortices are far apart and individually pinned, the vortex state is disordered. In weakly pinned samples showing the PE phenomenon, vortex solid is disordered in equilibrium above the peak position of PE. Modulations in S as a function of field between $H_{c1}$ and $H_p$ reflect the quality of spatial order in the intervening region. In particular the relative weights of the ordered/disordered pockets across the phase co-existence regime manifest themselves in a spectacular manner. An increase in the value of S between $H_{smp}^{p}$ and $H_p^{on}$ hints towards an improvement in the state of spatial order in this field interval due to possibility of squeezing out of some of the dislocations injected into the sample at the onset position of the SMP anomaly.

\section{Acknowledgment}

We would like to acknowledge numerous discussions with S. S. Banerjee, A. A. Tulapurkar and Ms. D. Jaiswal-Nagar. We would also like to thank Shobo Bhattacharya for his comments. We are grateful to C. V. Tomy for the crystal of Ca$_3$Rh$_4$Sn$_{13}$ grown at Warwick University, U.K. One of us (ADT) would like to acknowledge the TIFR Endowment Fund for the Kanwal Rekhi Career Development Fellowship.

\newpage

\begin{figure}

\includegraphics[scale=1.5,angle=0]{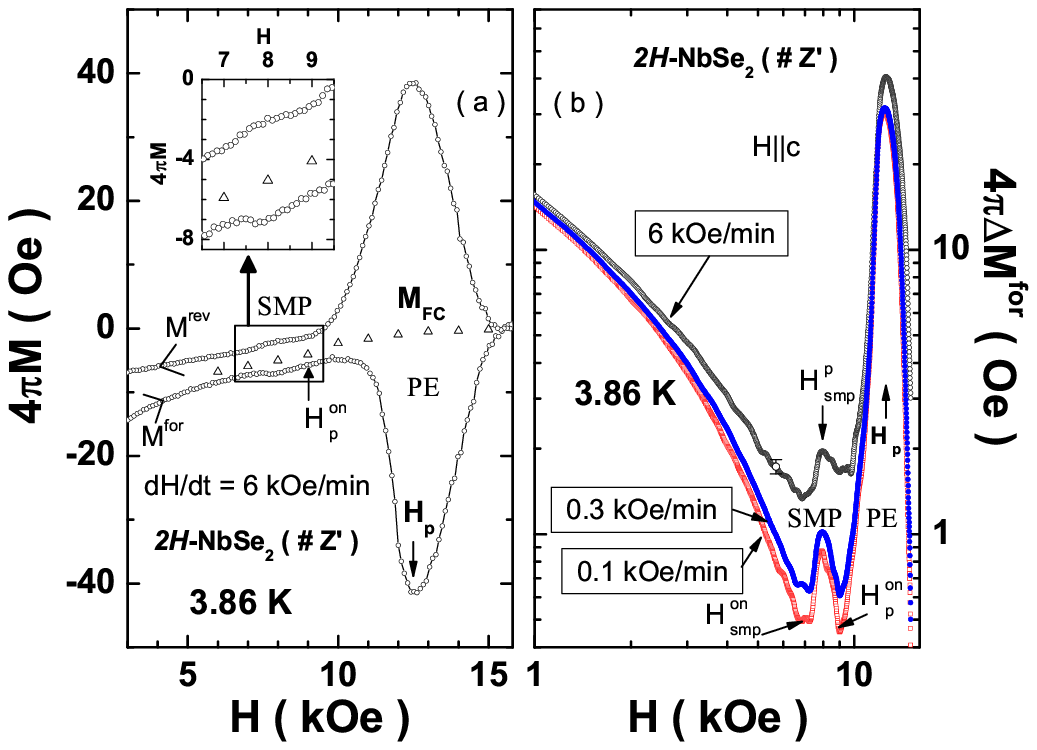}

\caption{(Color online)~Panel (a) shows a portion of the isothermal M-H loop ($H~\|~c$) at 3.86~K measured at a field scan rate of 6~kOe/min in the crystal $Z^{\prime}$ of $2H$-NbSe$_2$. The field cooled magnetization ($M_{FC}$) data points are also plotted as open triangles, and the PE and the SMP regions have been marked. Panel (b) shows plots of $\Delta M^{for}$ (obtained from the forward leg of the envelope M-H loop) at field scan rates of 0.1~kOe/min, 0.3~kOe/min and 6~kOe/min, respectively.}

\end{figure}

\begin{figure}

\includegraphics[scale=1.5,angle=0]{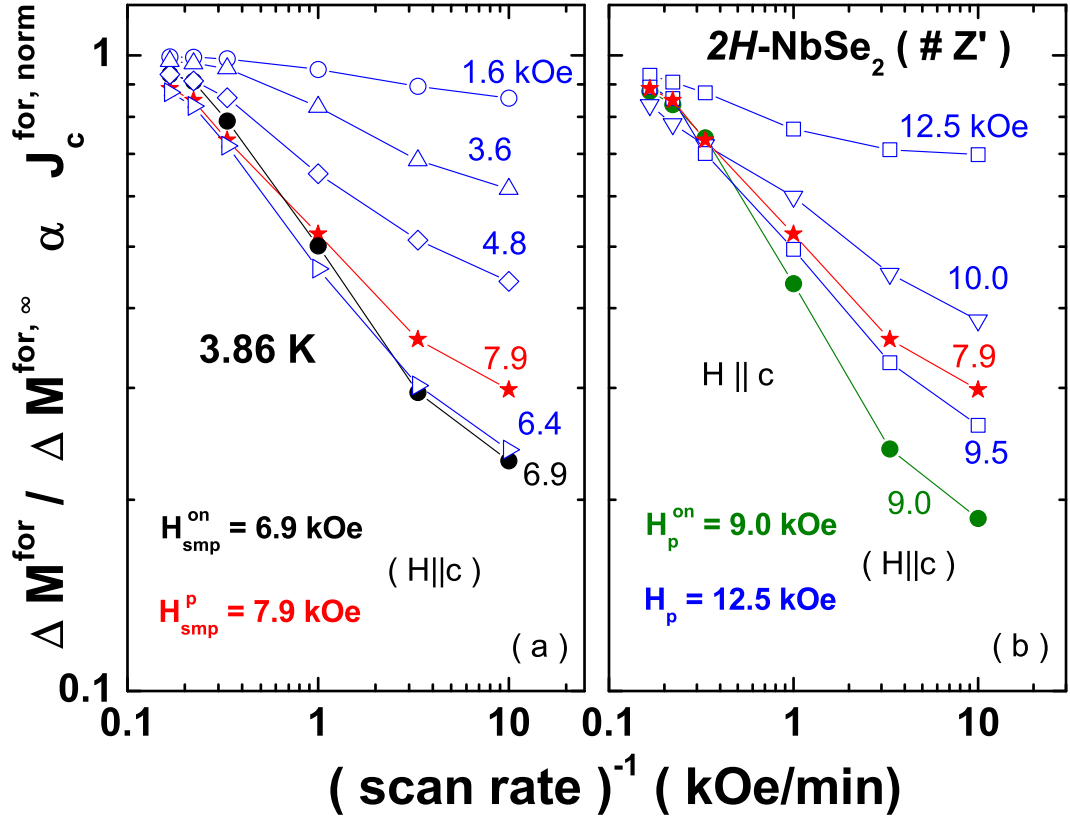}

\caption{(Color online)~Temporal variation of normalised critical current density $J_c^{for, norm}$ (see text for details) at various field values from deep within the mixed state right upto the peak position of the PE for the sample $Z^{\prime}$ of $2H$-NbSe$_2$ at 3.86~K.}

\end{figure}

\begin{figure}

\includegraphics[scale=1.5,angle=0]{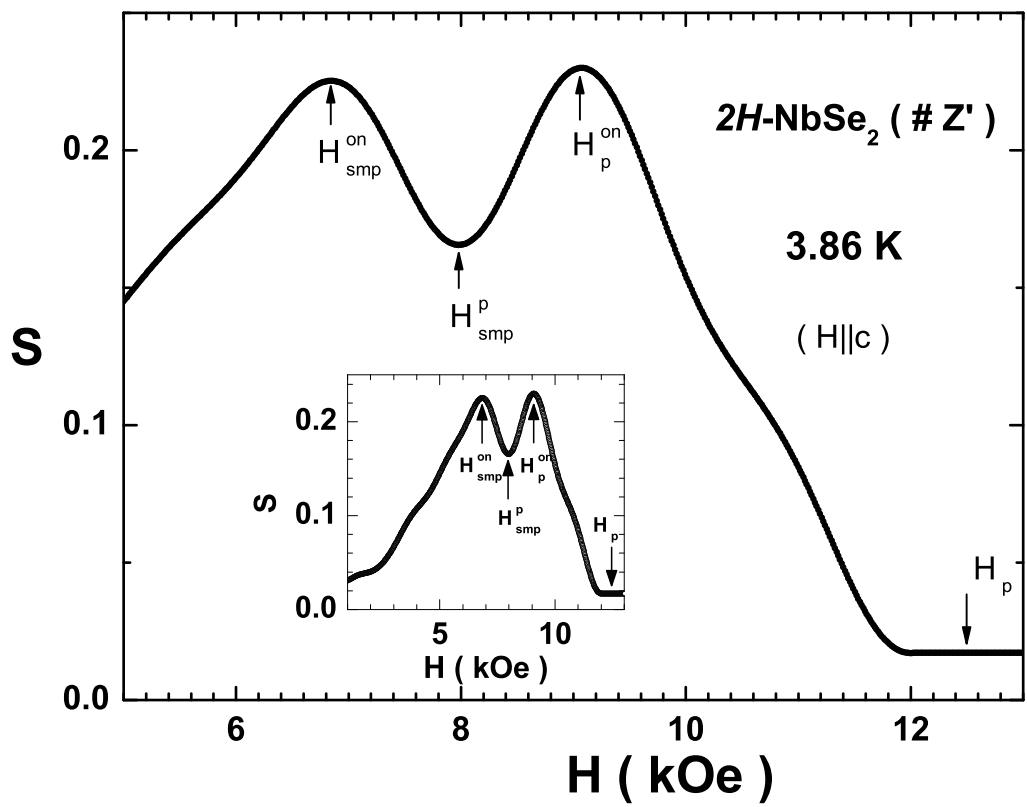}

\caption{$S$ parameter ($=\mid\frac{d~log(J_c^{for})}{d~log(\dot{H}^{-1})}\mid _{\dot{H}^{-1}~\rightarrow~\infty}$) as a function of field ($\|~c$) at 3.86~K in crystal $Z^{\prime}$ of $2H$-NbSe$_2$.}

\end{figure}

\begin{figure}

\includegraphics[scale=1.5,angle=0]{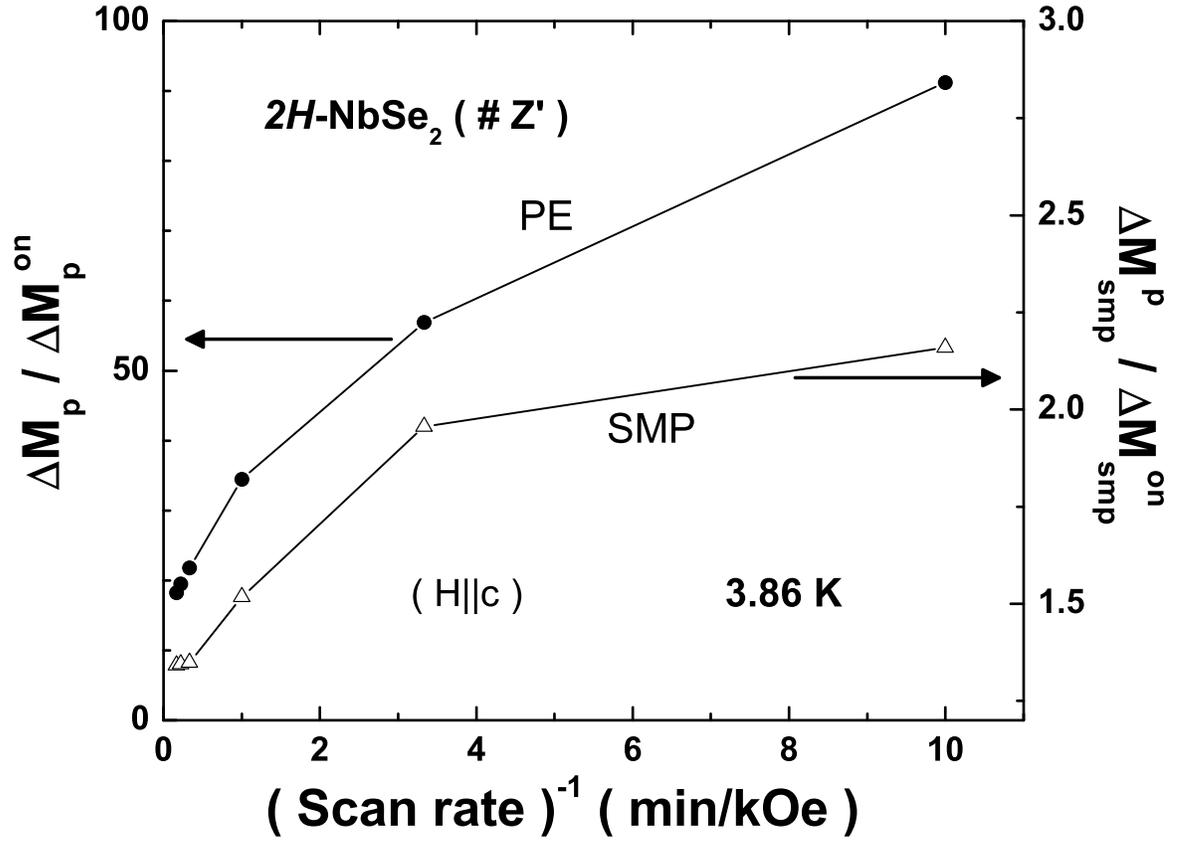}

\caption{Temporal variation of ratios of the hysteresis widths at the peak position to that at the the onset position for SMP and PE in the sample $Z^{\prime}$ of $2H$-NbSe$_2$ at 3.86~K.}

\end{figure}

\begin{figure}

\includegraphics[scale=1.5,angle=0]{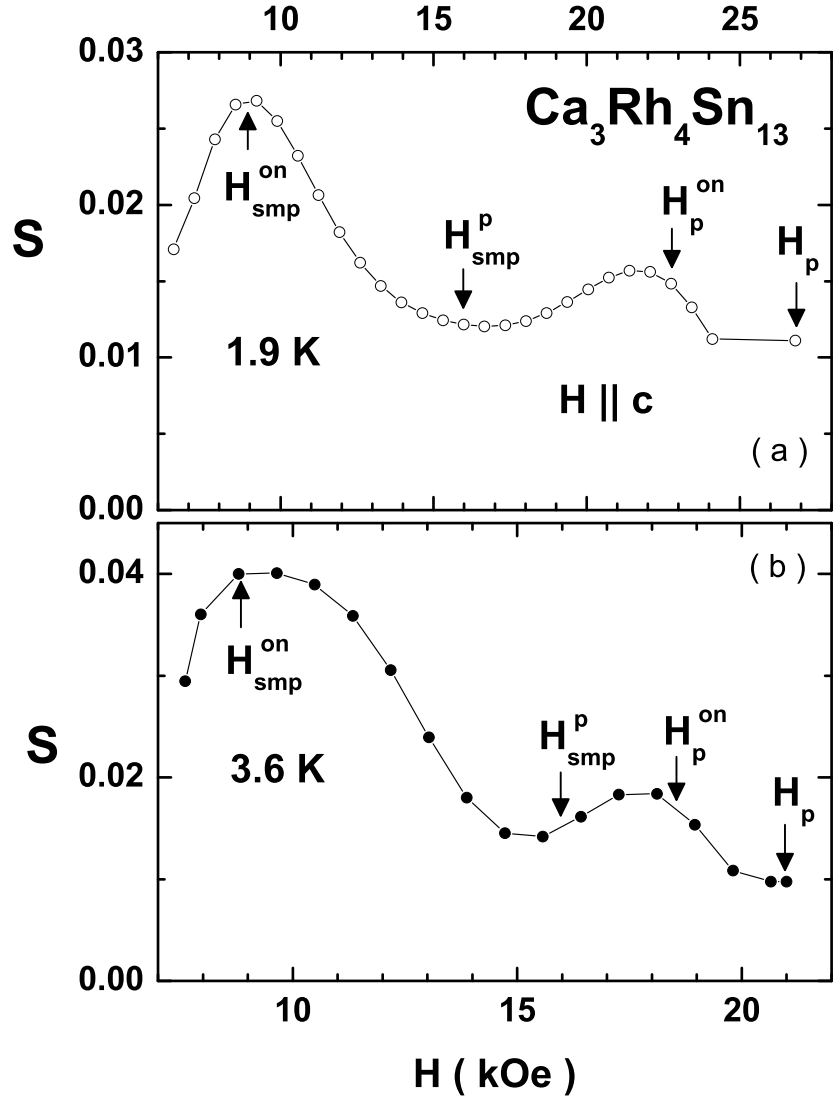}

\caption{S parameter as a function of field at 1.9~K and 3.6~K, respectively, in a weakly pinned crystal of Ca$_3$Rh$_4$Sn$_{13}$ ($T_c(0) \approx 8.2~K$).}

\end{figure}

\begin{figure}

\includegraphics[scale=1.5,angle=0]{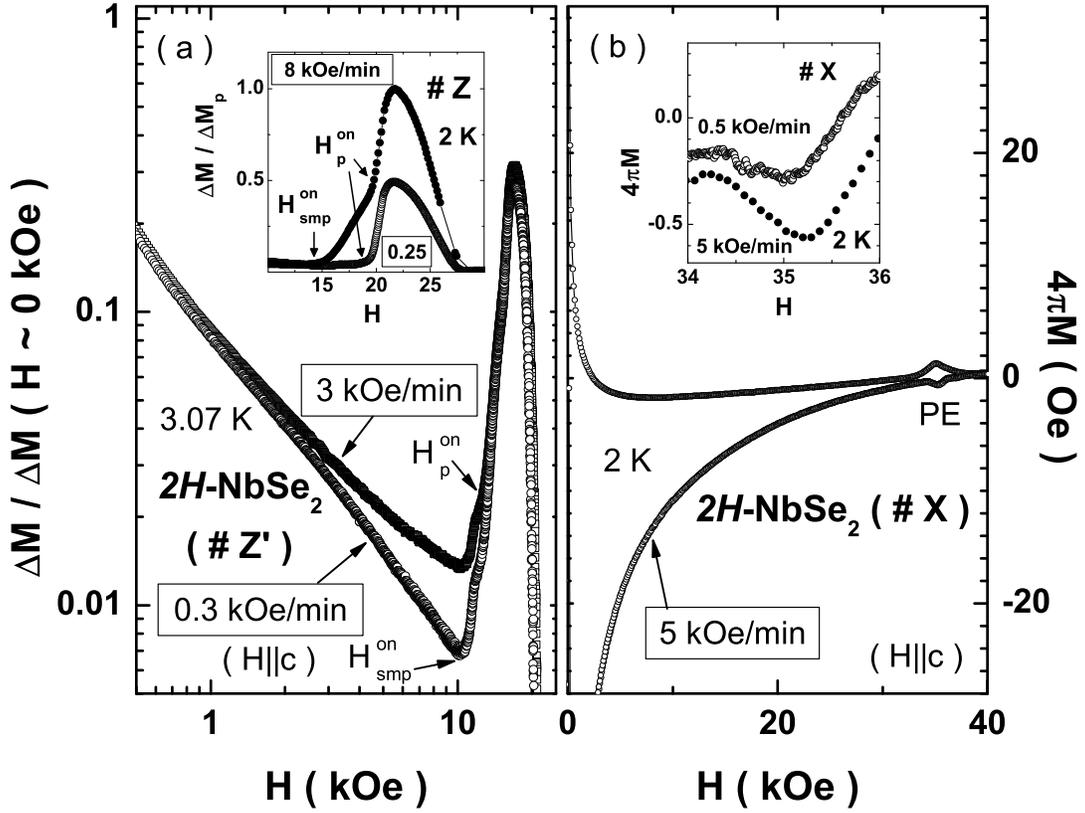}

\caption{Panel (a) shows a plot of $\Delta M^{for}/ \Delta M^{for}(H \sim 0~kOe)$ as a function of field at scan rates of 0.3~kOe/min and 3~kOe/min for the sample $Z^{\prime}$ of $2H$-NbSe$_2$ at 3.07~K. An inset in panel (a) shows the magnetization hysteresis width in the sample $Z$ of $2H$-NbSe$_2$ at 2~K as the scan rate is varied from 8~kOe/min to 0.25~kOe/min. The notional positions of onset of SMP and PE anomalies have been indicated. Panel (b) shows a portion of the two quadrant isothermal M-H loop in sample $X$ of $2H$-NbSe$_2$ at 2~K, measured with a scan rate of 5~kOe/min. An inset in panel (b) shows a comparison of the forward legs of the M-H loop in the PE region measured at scan rates of 5~kOe/min and 0.5~kOe/min, respectively.}

\end{figure}

\begin{figure}

\includegraphics[scale=1.5,angle=0]{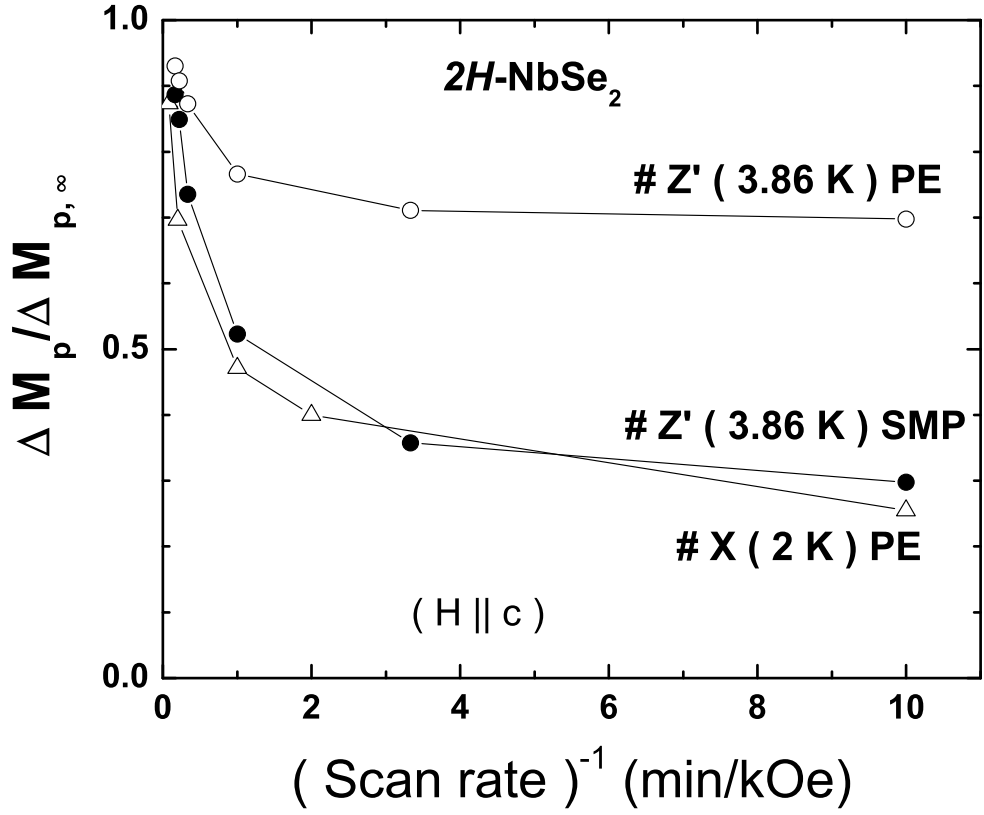}

\caption{Temporal variation of $\Delta M^{for}/ \Delta M^{for, \infty}$ at the peak positions of PE and SMP for the sample $Z^{\prime}$  and that corresponding to peak position of the PE  for sample $X$.}

\end{figure}

\begin{figure}

\includegraphics[scale=1.5,angle=0]{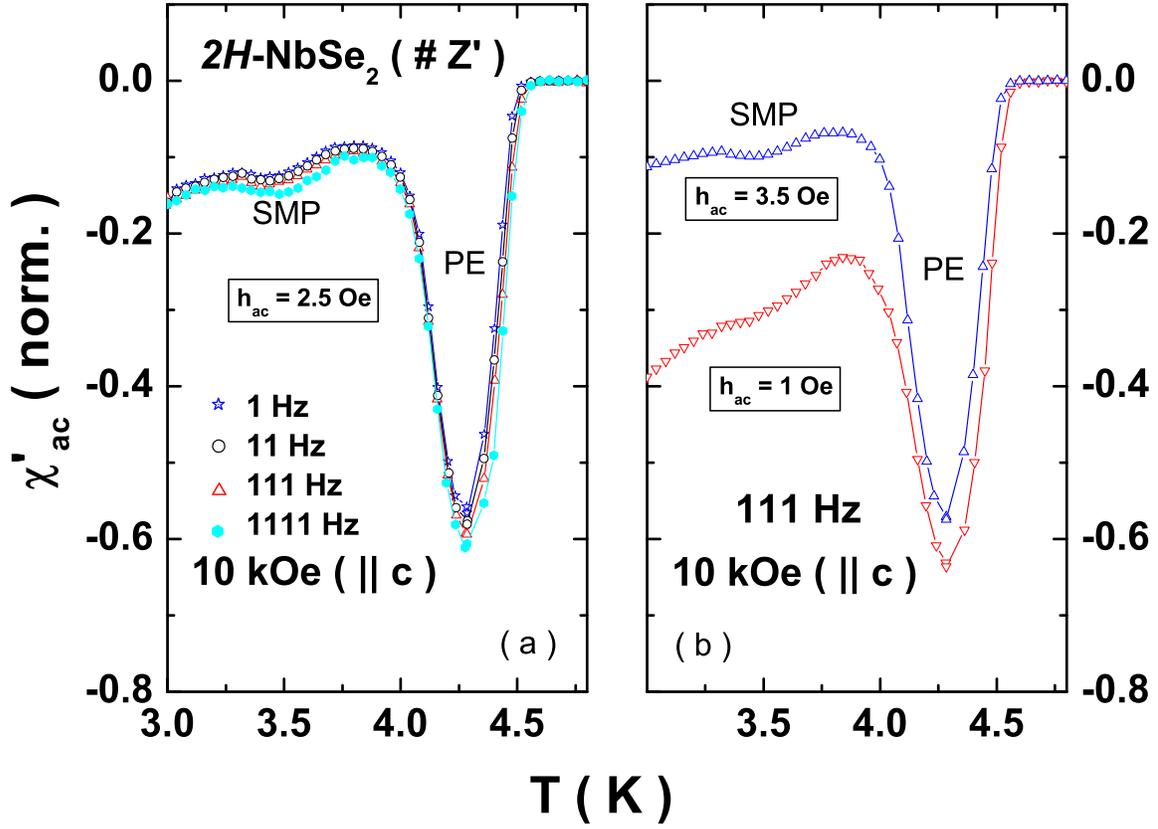}

\caption{(Color online)~Panel (a) shows ac susceptibility ($\chi^{\prime}_{ac}(T)$) data in dc magnetic field of 10~kOe, measured with $h_{ac}$ of 2.5~Oe (r.m.s.) in the frequency range 1~Hz to 1111~Hz. Panel (b) shows the $\chi^{\prime}_{ac}(T)$ data measured at two amplitudes of the ac drive ($h_{ac}$ of 1.0~Oe (r.m.s.) and 3.5~Oe (r.m.s.)) at a measuring frequency of 111~Hz.}

\end{figure}

\begin{figure}

\includegraphics[scale=1.5,angle=0]{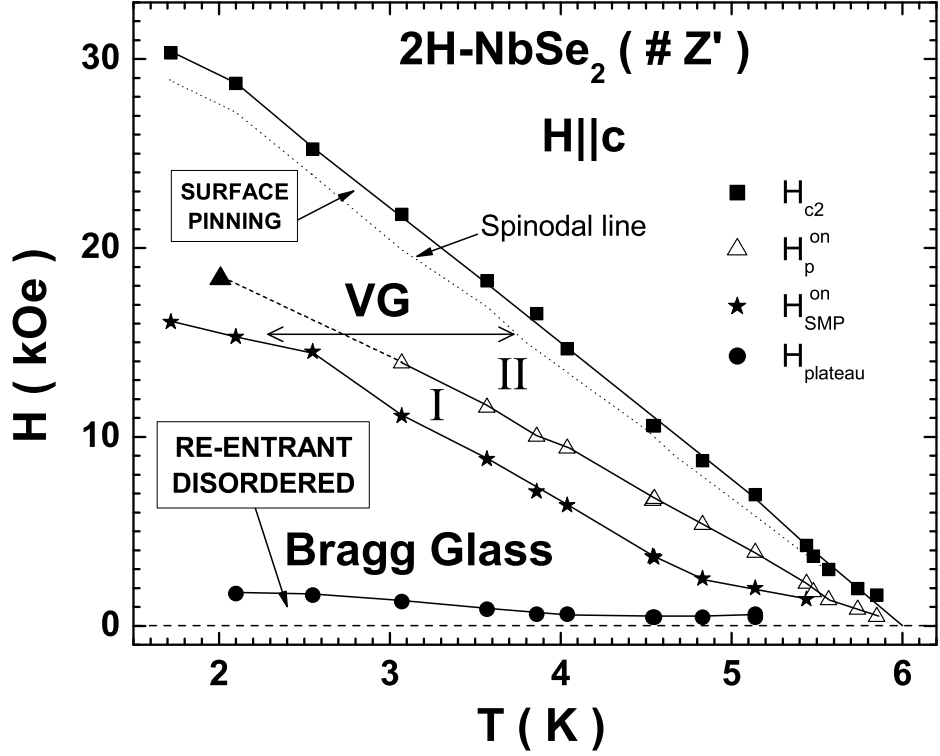}

\caption{Vortex phase diagram constructed from isothermal M-H measurements done at a high scan rate (viz., 3~kOe/min) in the crystal $Z^{\prime}$ of $2H$-NbSe$_2$, for $H~\|~c$. The re-entrant disordered phase, Bragg glass phase, VG phase (sub-parts I and II) and the surface pinning regions have been marked. The spinodal line as in Ref. \cite{adtprb} has also been drawn.}

\end{figure}

\end{document}